\documentclass[10pt, conference, compsocconf]{IEEEtran}

\pdfoutput=1

\usepackage{cite}
\usepackage{amsmath,amssymb,amsfonts}
\usepackage{algorithmic}
\usepackage{graphicx}
\usepackage{textcomp}
\usepackage{bm}
\usepackage{xcolor}
\def\BibTeX{{\rm B\kern-.05em{\sc i\kern-.025em b}\kern-.08em
    T\kern-.1667em\lower.7ex\hbox{E}\kern-.125emX}}
    
\usepackage[font=footnotesize]{subfig}
\usepackage{color}

\usepackage{mathtools}
\DeclarePairedDelimiter{\ceil}{\lceil}{\rceil}

\newcommand{\etal}{\textit{et al}.}

\begin{document}

\title{Learning with Analytical Models}

\author{\IEEEauthorblockN{Huda Ibeid, Siping Meng, Oliver Dobon, Luke Olson, William Gropp}
\IEEEauthorblockA{University of Illinois at Urbana-Champaign\\
Urbana, IL, USA\\
Email: \{hibeid,smeng10,dobon2,lukeo,wgropp\}@illinois.edu}
}

\maketitle

\begin{abstract}
To understand and predict the performance of scientific applications, several analytical and machine learning approaches have been proposed, each having its advantages and disadvantages. In this paper, we propose and validate a hybrid approach for performance modeling and prediction, which combines analytical and machine learning models. The proposed hybrid model aims to minimize prediction cost while providing reasonable prediction accuracy. Our validation results show that the hybrid model is able to learn and correct the analytical models to better match the actual performance. Furthermore, the proposed hybrid model improves the prediction accuracy in comparison to pure machine learning techniques while using small training datasets, thus making it suitable for hardware and workload changes.

\end{abstract}


\IEEEpeerreviewmaketitle

\section{Introduction}

Performance modeling has been extensively used to understand and predict the performance of scientific applications. However, the increasing complexity of modern computing architectures along with the exponentially growing configuration space and complex interactions among configuration options often make it difficult to develop accurate performance models.

Classical approaches for performance modeling rely on two techniques, analytical modeling (AM) and machine learning (ML). The basic idea behind analytical models is to represent application program by means of a set of analytical equations. The analytical models typically rely on simplifying assumptions about the behavior of the underlying architecture and application. Thus, the analytical model accuracy can be challenged when these assumptions are not matched. Machine learning, on the other hand, relies on observing the actual performance in order to infer statistical models. The accuracy of machine learning performance models depends on the representativeness of the training dataset. Unfortunately, the space of all possible configurations grows exponentially with the number of variables (the curse of dimensionality).

In this paper, we propose a hybrid approach for performance modeling and prediction, which couples analytical and machine learning models. Our aim is to combine the evaluation speed of analytical models with the architecture awareness of machine learning models in order to achieve a model that carries out predictions with reasonable accuracy without too many time consuming and costly experiments for data collection.

The proposed hybrid model is conceptually simple and easy to implement. It consists of analytical models of the corresponding application code, two ensemble methods, a training algorithm, and a prediction algorithm. In particular, our proposed hybrid model uses stacking and bagging ensemble methods, stacking improves predictions while bagging reduces variance and helps to prevent overfitting~\cite{wolpert1992, breiman1996}. We illustrate and validate our approach using two applications, a 7-point 3-D stencil code from the PATUS DSL source-to-source stencil compiler~\cite{Christen2011}, which generates C code with multi-threading and SIMD instructions, and a fast multipole method code, ExaFMM~\cite{yokota2013fmm}, which support hybrid MPI/OpenMP parallelism.

This paper is organized as follows. Section~\ref{sec:applications} briefly describes stencil computation and fast multipole method. Section~\ref{sec:setup} presents the experimental setup. Analytical models for stencil computation and FMM are discussed in Section~\ref{sec:AM}. The machine learning methodology is described in Section~\ref{sec:ML}. Section~\ref{sec:hybrid} presents our hybrid model. Section~\ref{sec:evaluation} evaluates the hybrid model and discusses the experimental results. Section~\ref{sec:related} elaborates on related work on modeling. Finally, Section~\ref{sec:conclusion} contains conclusions.

\section{Applications}
\label{sec:applications}

\subsection{Stencil Computation}
\label{subsec:stencil}

\begin{figure}[t]
    \centering
     \includegraphics[width=0.5\linewidth]{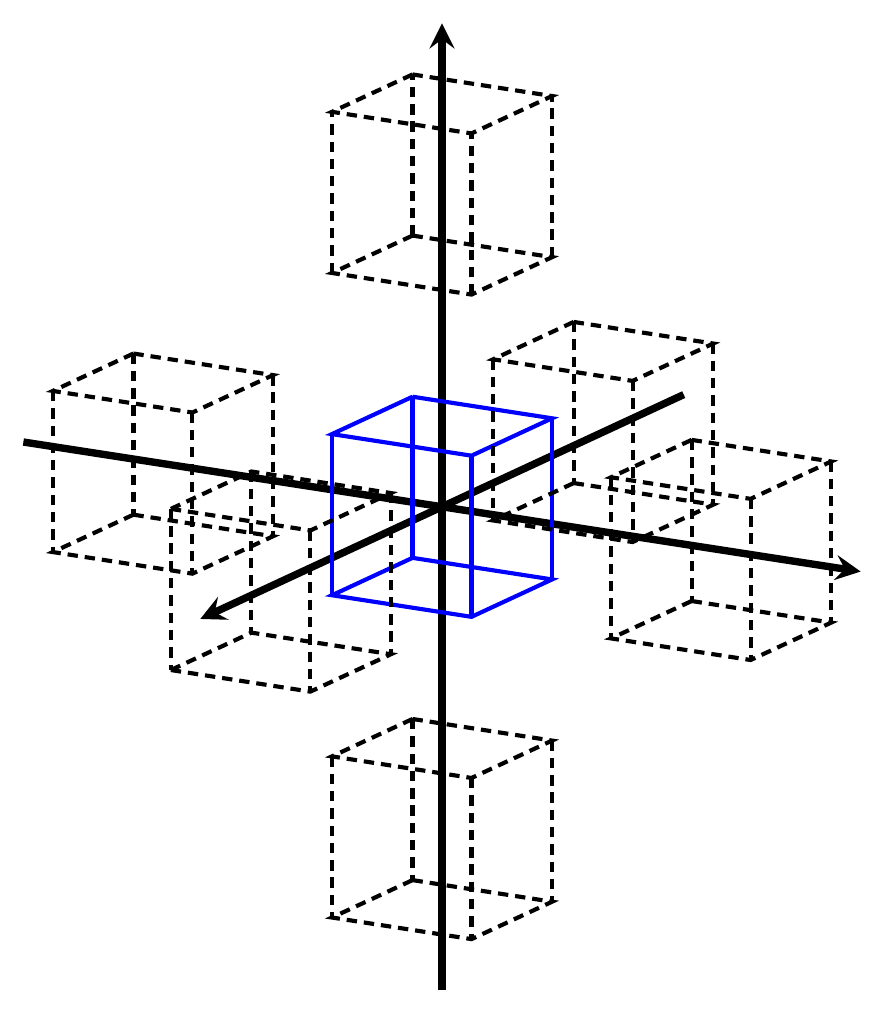}
    \caption{Schematic diagram of the 7-point 3-D stencil.}
    \label{fig:stencilDiagram}
\end{figure}

Partial differential equations (PDEs) arise in a vast number of applications in scientific computing in diverse areas such as heat diffusion, electromagnetics, and fluid dynamics. A common method for solving PDEs on regular grids is to discretize them by finite-difference techniques and then solve the resulting large, sparse linear systems. Finite-difference iterative algorithms apply the same update operation to all grid points at each time step. This operation requires access to a fixed neighborhood of the grid point and is thus denoted as stencil operation. In a stencil computation, each point of the computational domain is updated with weighted contributions from a subset of its neighbors in both time and space. Depending on the number of neighbors contributing, including the point itself, an x-point stencil is formed. A 7-point or a 27-point stencil is often used for 3-D domains. Figure~\ref{fig:stencilDiagram} shows an illustration of the 7-point stencil where the central point is updated by the weighted average of six of its neighbors. In general, stencils represent computational patterns that repeat across the computational domain. These patterns are then used to build solvers that range from simple Jacobi iterations to complex multigrid and block structured adaptive methods~\cite{Datta2009}. Below is a pseudocode for the 7-point 3-D classical stencil algorithm
\begin{algorithmic}
\FOR {$t \leftarrow 0$ \TO $timesteps$}
  \FOR{$k \leftarrow 1$ \TO $KK - 1$}
    \FOR{$j\leftarrow 1$ \TO $JJ - 1$}
      \FOR{$i\leftarrow 1$ \TO $II - 1$}
        \STATE$\chi^t_{i,j,k} = C_0 \times \chi^{t - 1}_{i,j,k} + C_1 \times (\chi^{t - 1}_{i-1,j,k} + \chi^{t - 1}_{i+1,j,k} + \chi^{t - 1}_{i,j-1,k} + \chi^{t - 1}_{i,j+1,k} + \chi^{t - 1}_{i,j,k-1} + \chi^{t - 1}_{i,j,k+1})$\;
      \ENDFOR
    \ENDFOR  
  \ENDFOR
\ENDFOR
\label{algo:stencil}
\end{algorithmic}
where $II$, $JJ$, and $KK$ are the grid dimensions including ghost points and $C_0$ and $C_1$ are the spatial discretization coefficients.

Stencil operations typically perform a very limited number of operations per grid point, and data from main memory can not be transferred fast enough to avoid stalling computations. Hence, stencil computations performance is bounded by the available memory bandwidth and only achieves a small fraction of the theoretical peak performance. It is therefore crucial to optimize stencil computations in order to improve applications performance and reduce execution time. Various optimization techniques have traditionally been applied to stencil computations such as diamond tiling~\cite{Bertolacci2015}, multi-dimensional tiling~\cite{malas2016}, time skewing~\cite{Strzodka2011}, SIMD instructions~\cite{Henretty2011}, and multi-threading~\cite{kamil2010}. However, as computer architectures are becoming increasingly complex, writing efficient scientific codes that make best use of the available resources is becoming more difficult. The complex interaction between application-level optimizations and the underlying architecture makes it difficult to find an optimal set of optimizations. Additionally, the performance achieved by specific set of optimizations is usually not portable between different architectures or different stencil codes. In this paper, we demonstrate that hybrid models obtained by coupling analytical models with machine learning techniques are able to accurately predict performance in the regimes of interest while decreasing the cost of modeling across environments.

\subsection{Fast Multipole Method}
\label{subsec:fmm}

$N$-body problems are used to simulate physical systems of particles interaction under physical or electromagnetic field~\cite{greengard1987}. The $N$-body problem can be represented by the sum
\begin{equation}
f(y_j) = \sum_{i=1}^N w_i K(y_j,x_i),
\end{equation}
where $f(y_j)$ represents a field value evaluated at a point $y_j$ which is generated by the influence of sources located at $\{x_i\}$ with weights $w_i$. $K(y_j,x_i)$ is the kernel that governs the interactions between evaluation and source particles.

The direct approach to simulate the $N$-body problem evaluates all pair-wise interactions among the particles which results in a computational complexity of $\mathcal{O}(N^2)$. This complexity is prohibitively expensive even for modestly large datasets. For simulations with large datasets, many faster algorithms have been invented, such as the fast multipole methods (FMM)~\cite{greengard1987}. FMM divides the computational domain into near-domain and far-domain and computes interactions between clusters by means of local and multipole expansions, providing $\mathcal{O}(N)$ complexity. FMM is more than an $N$-body solver, however. Recent efforts to view the FMM as an elliptic PDE solver have opened the possibility to use it as a preconditioner for even a broader range of applications~\cite{Ibeid2018}.

The first step of the FMM algorithm is the decomposition of the computational domain. This spatial decomposition is accomplished by a hierarchical subdivision of the space associated with a tree structure. The 3-D spatial domain of FMM is represented by oct-trees, where the space is recursively subdivided into eight cells until the finest level of refinement or ``leaf level''.

\begin{figure}
\centering
\includegraphics[width=\linewidth]{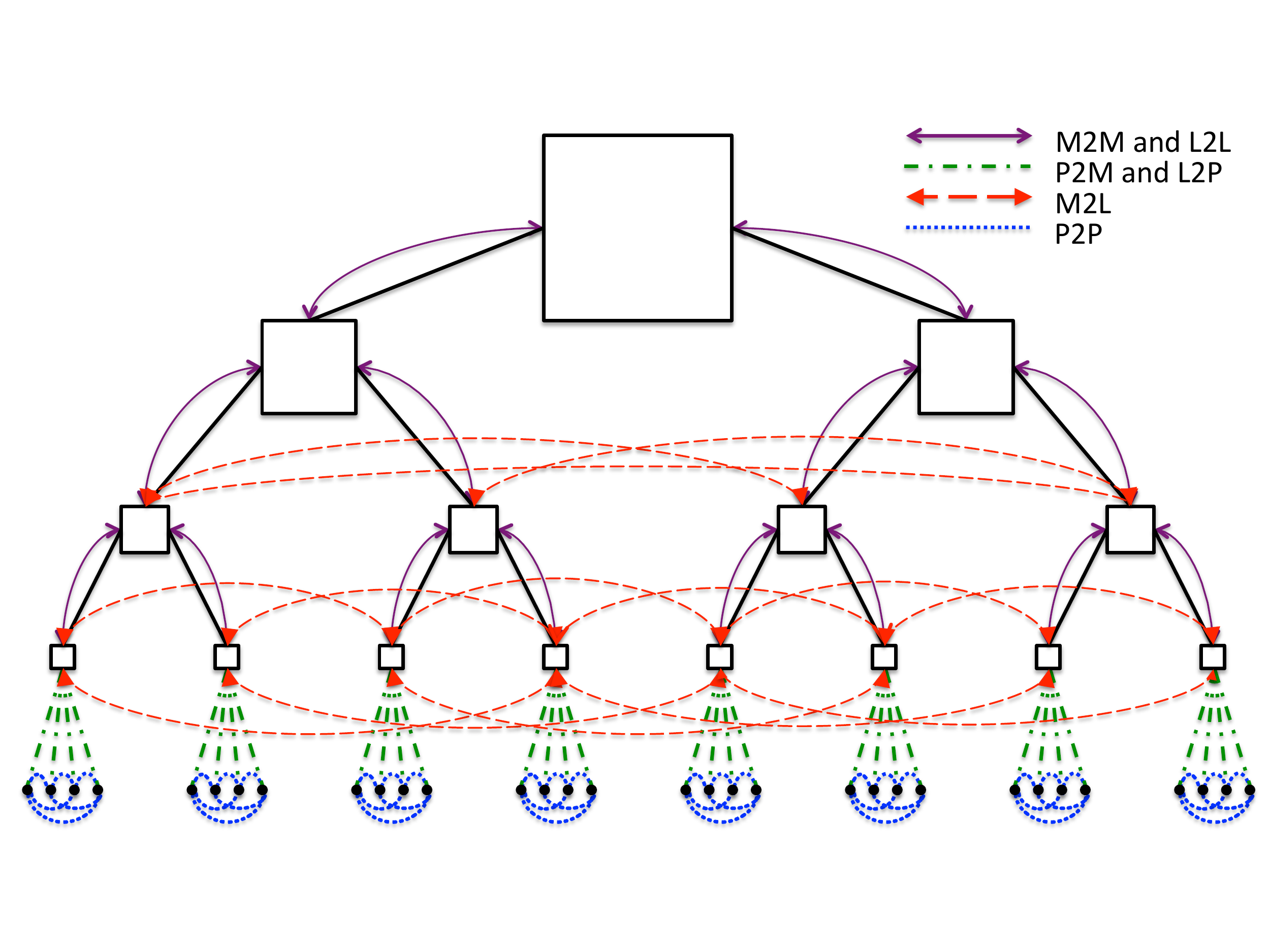}
\caption{Illustration of the FMM kernels: P2M (Particle-to-Multipole), M2M (Multipole-to-Multipole), M2L (Multipole-to-Local), L2L (Local-to-Local), L2P (Local-to-Particle), and P2P (Particle-to-Particle).}\label{fig:fmm_flow}
\end{figure}

The FMM calculation begins by transforming the mass/charge of the source particles into multipole expansions by means of a Particle-to-Multipole kernel (P2M). Then, the multipole expansions are translated to the center of larger cells using a Multipole-to-Multipole kernel (M2M). FMM calculates the influence of the multipoles on the target particles in three steps: (1) translation of the multipole expansions to local expansions between well-separated cells using a Multipole-to-Local kernel (M2L); (2) translation of local expansions to smaller cells using a Local-to-Local kernel (L2L); and (3) translation of the effect of local expansions in the far field onto target particles using a Local-to-Particle kernel (L2P). All-pairs interaction is used to calculate the near field influence on the target particles by means of a Particle-to-Particle kernel (P2P). Figure~\ref{fig:fmm_flow} illustrates the FMM main kernels: Particle-to-Multipole (P2M), Multipole-to-Multipole (M2M), Multipole-to-Local (M2L), Local-to-Local (L2L), Local-to-Particle (L2P), and Particle-to-Particle (P2P). The dominant kernels of the FMM calculation are P2P and M2L.

\section{Experimental Setup}
\label{sec:setup}

\subsection{Machine Description}

We ran our experiments on the Blue Waters supercomputer which is a Cray XE6/XK7 system managed by the National Center for Supercomputing Applications for the National Science Foundation. Though Blue Waters contains XE6 and XK7 nodes, our experiments were confined to the XE nodes. The XE6 dual-socket nodes are populated with 2 AMD Interlagos model 6276 CPU processors (one per socket) with a nominal clock speed of at least 2.3 GHz and 64 GB of physical memory. Each Interlagos processor is composed of eight Bulldozer cores where each core has 16KB/64KB data/instruction write-through L1 cache, 2 MB write-back L2 cache, and an 8 MB shared write-back L3 cache.

\subsection{Benchmarks and Scientific Applications}

Different frameworks include different code optimizations for stencil applications. In this paper, we use PATUS~\cite{Christen2011} to validate our stencil models. PATUS is a code generation framework for stencil computations targeted at modern multi- and many-core processors using domain specific language (DSL). PATUS exposes loop blocking, loop unrolling, and multi-threading. Loop blocking is applied to all loop levels requiring three blocking sizes ($b_i$, $b_j$, and $b_k$) to tune. Afterwards, loop unrolling ($u$) is applied to the innermost loop. The unrolling factor may vary between 0 (no unrolling) and 8. Therefore, our PATUS modeling vector $\bm{X} = (I, J, K, b_i, b_j, b_k, u, t)$ where $I$, $J$, and $K$ are the grid dimensions and $t$ is the number of threads.

ExaFMM~\cite{yokota2013fmm} is an open source library for fast multipole methods aimed towards Exascale systems. It provides error aware local optimizations, a multipole acceptance criterion, and employs dual tree traversal which is an efficient strategy for finding the list of cell-cell interactions. ExaFMM is highly scalable to millions of cores with support for MPI, thread-level, and SIMD parallelism. In this paper, we use the Laplace kernel in three dimensions with random distribution of particles in a cube. Our ExaFMM modeling vector $\bm{X} = (t, N, q, k)$ where $t$ is the number of threads, $N$ is the total number of particles, $q$ is the number of particles per leaf cell, and $k$ is the order of expansion.

\section{Analytical Modeling}
\label{sec:AM}

To model computational cost, we must consider the costs of both floating-point operations ($T_{flops}$) as well as memory operations ($T_{mem}$). These two costs combined reflect the single-node performance which is a critical building-block in scalable parallel codes. Assuming arithmetic and memory operations can be overlapped, the time $T$ required to solve a problem of size $N$ can be approximated by
\begin{equation}
T = \max \big(T_{flops} , T_{mem} \big),
\end{equation}
where $T_{flops}$ is defined as the total number of floating-point operations, multiplied by the time per floating-point operation ($t_c$) in seconds and $T_{mem}$ is modeled as the total data fetched into fast memory, multiplied by the memory bandwidth inverse ($\beta_{\textnormal{mem}}$) in units of seconds per element.

\subsection{Stencil Computation}

To analytically model the 3-D stencil computations, we start with the models introcudced in~\cite{de2011}. As stencil computations are memory-bound, the cost of computing the floating-point operations is assumed negligible due to the overlap with memory transfers.

Given a grid of size $N = I \times J \times K$ elements of order $l$, where $I$, $J$, and $K$ correspond to the $x$, $y$, and $z$ dimensions, respectively, the total memory requirement to compute a Y-X plane (one k iteration) is given by
\begin{equation}
S_{total} = P_{read} \times S_{read} + P_{write} \times S_{write},
\label{eq:s_wb}
\end{equation}
where $P_{read} = 2 \times l + 1$, $P_{write} = 1$, $S_{read} = II \times JJ$, and $S_{write} = I \times J$ where $P$ and $S$ denote the number of planes and plane size, respectively, and $II$, $JJ$, and $KK$ represent the dimensions of the problem including ghost points. This model assumes a write-allocate cache. For a no-write-allocate cache,~\eqref{eq:s_wb} can be rewritten as
\begin{equation}
S_{total} = P_{read} \times S_{read}.
\label{eq:s_wt}
\end{equation}

On an architecture with a memory hierarchy of $n$ cache levels, the total time to compute a stencil is
\begin{equation}
T = T_{L1} + T_{Li} + \dots + T_{Ln} + T_{mem},
\end{equation}
where $T_{Li}$ and $T_{mem}$ are the time to access data in $Li$ cache level and main memory, respectively.

In general, the time spent moving data for $Li$ cache level is given by
\begin{equation}
T_{Li} = T^{data}_{Li} \times Hits_{Li},
\end{equation}
where $T^{data}_{Li} = data * \beta_{\textnormal{mem}_{Li}}$ is the time required to transfer data (element or cacheline) from level $i$ and $Hits_{Li} = Misses_{Li - 1} - Misses_{Li}$ is the number of transfers performed. The amount of misses issued at each cache level is estimated by
\begin{equation}
Misses_{Li} = \lceil II / W \rceil \times JJ \times KK \times nplanes_{Li},
\label{eq:misses}
\end{equation}
where $W$ is the number of elements per cacheline and $nplanes_{Li}$ is the number of $II \times JJ$ planes read from the next cache level for each $k$ iteration due to possible compulsory, conflict, or capacity misses. The computation of $nplanes_{Li}$ is yield by the following conditional equations
\[
    nplanes_{Li} = 
\begin{cases}
    1 ,& \text{if } R_1\\
    (1, P_{read} - 1], & \text{if } \neg R_1 \land R_2\\
    (P_{read} - 1, P_{read}], & \text{if } \neg R_2 \land R_3\\
    (P_{read}, 2 \times P_{read} - 1],  & \text{if } \neg R_3 \land \neg R_4\\
    2 \times P_{read} - 1, & \text{if } R_4
\end{cases}
\]
where 
\begin{align*}
R_1 &: ((size_{Li}/W) \times R_{col} \geq S_{total}),\\
R_2 &: ((size_{Li}/W) > S_{total}),\\
R_3 &: ((size_{Li}/W) \times R_{col} > S_{read}).\\
R_4 &: ((size_{Li}/W) \times R_{col} < P_{read} \times II),\\
\end{align*}
where $R_{col} = P_{read} / (2 \times P_{read } - 1)$ and $size_{Li}$ is the cache size. We use linear interpolation to smooth discontinuities that appears when transitioning from one case to another.

\subsection{Fast Multipole Method}

In this section, we present analytical models for the two phases of FMM that consume most of the calculation time\@: P2P and M2L\@. We assume a nearly uniform particles distribution and therefore a full oct-tree structure.

\subsubsection{Computation Costs}

\paragraph{\bf P2P}
Assuming $q$ particles per leaf cell, the computational complexity of the P2P phase is $27 q^2 \frac{N}{q}$. This leads to a computation cost of
\begin{equation}
T_{\textnormal{flop},\textnormal{P2P}} = 27 \cdot q N \cdot t_c.
\end{equation}
where $N$ is the problem size.

\paragraph{\bf M2L} The asymptotic complexity of the M2L phase depends on the order of expansion $k$ and the choice of series expansion. ExaFMM uses Cartesian series expansion which has operations count of $189 k^{6}$. Hence
 \begin{equation}
 T_{\textnormal{flop},\textnormal{M2L}} = 189 \cdot \frac{N k^6}{q} \cdot t_c.
 \end{equation}

\subsubsection{Memory Access Costs}

As shown in~\cite{Chandramowlishwaran2012}, the outer loops of the P2P and M2L computations can be modeled as sparse matrix-vector multiplies. For a cache with size $Z$ and cache-line length $L$ in elements, a cache-oblivious algorithm~\cite{Blelloch2010} for multiplying a sparse $H \times H$ matrix with $h$ non-zeros by a vector establishes an upper bound on cache misses in the SpMV as
\begin{equation}\label{eq:SpMV}
\mathcal{O}(\frac{h}{L}+\frac{H}{Z^{1/3}}),
\end{equation}
for each transferring line of size $L$.

\paragraph{\bf P2P}
Applying~\eqref{eq:SpMV} gives an upper bound on the number of cache misses for the P2P phase as follows
\begin{equation}\label{eq:P2P_mem}
Q_{\textnormal{P2P}} \leq 4 \cdot \frac{N}{L}+ b_{\!_{\textnormal{P2P}}} \cdot \frac{N/q}{L}+ 4 \cdot \frac{N}{L}+\frac{N/q}{{(\frac{Z}{4q})}^{\!^{1/3}}},
\end{equation}
where $b_{\!_{\textnormal{P2P}}}$ is the average number of source cells in the neighbor list of a target leaf cell ($b_{\!_{\textnormal{P2P}}} = 26$ for an interior cell in a uniform distribution). The first two terms on the right-hand side of~\eqref{eq:P2P_mem} refer to read accesses for the source cells and the neighbor lists for each target cell, while the third term refers to the update accesses for the target leaf cell potentials. In P2P communication, coordinates and values of every particle belonging to the cell must be sent, resulting in a multiplication factor of four. We model the dominant access time as
\begin{equation}
T_{\textnormal{mem},\textnormal{P2P}} = N \cdot \beta_{\textnormal{mem}} + \frac{NL}{(Z^{(1/3)}q^{(2/3)})} \cdot \beta_{\textnormal{mem}}.
\end{equation}

\paragraph{\bf M2L}
Applying~\eqref{eq:SpMV} for the M2L phase gives an upper bound on the number of cache misses as follows
\begin{equation}\label{eq:M2L_mem}
Q_{\textnormal{M2L}} \leq \frac{(b_t+bs) f(k)}{L}+\frac{b_{\!_{\textnormal{M2L}}}b_t}{L}+\frac{b_t}{{\big(\frac{Z}{f(p)}\big)}^{\!^{1/3}}},
\end{equation}
where $b_t$ is the number of target cells, $b_s$ is the number of source cells, $b_{\!_{\textnormal{M2L}}}$ is the average number of source cells in the well-separated list of a target cell ($b_{\!_{\textnormal{M2L}}} = 189$ for an interior cell in a uniform distribution), and $f(k)$ is the asymptotic complexity.

Considering the higher order terms, the memory access cost of the M2L phase can be approximated by
\begin{equation}
T_{\textnormal{mem},\textnormal{M2L}} = \frac{N  k^{6}}{q} \cdot \beta_{\textnormal{mem}} + \frac{N { k^{2}} L}{q Z^{1/3}} \cdot \beta_{\textnormal{mem}}.
\end{equation}
\hfill

\section{Machine Learning}
\label{sec:ML}

\begin{figure*}
\centering
\begin{minipage}{1\linewidth}
\centering
	\subfloat[Decision Trees]{\includegraphics[width=0.3\textwidth]{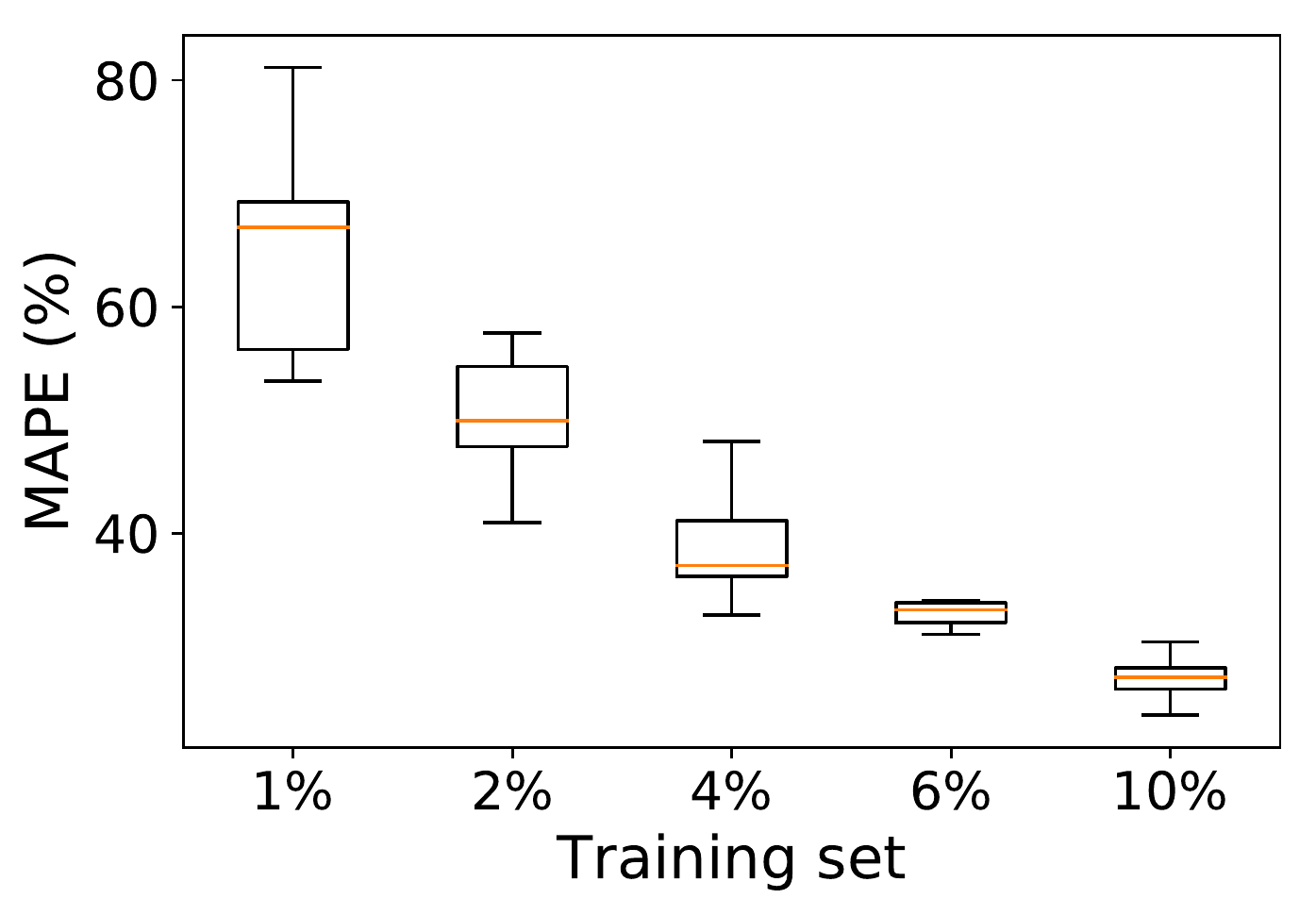}}
	\subfloat[Extra Trees]{\includegraphics[width=0.3\textwidth]{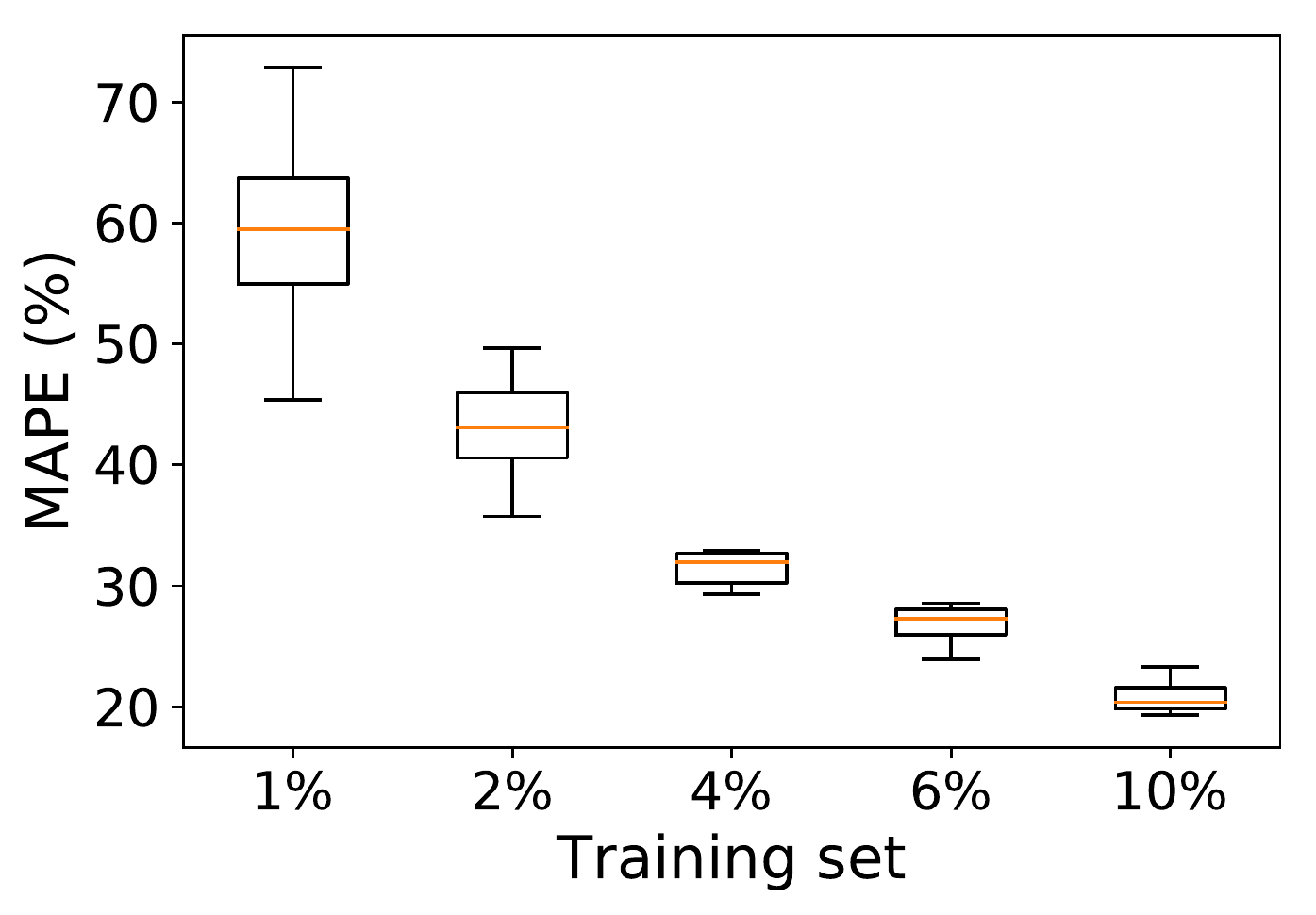}}
	\subfloat[Random Forests]{\includegraphics[width=0.3\textwidth]{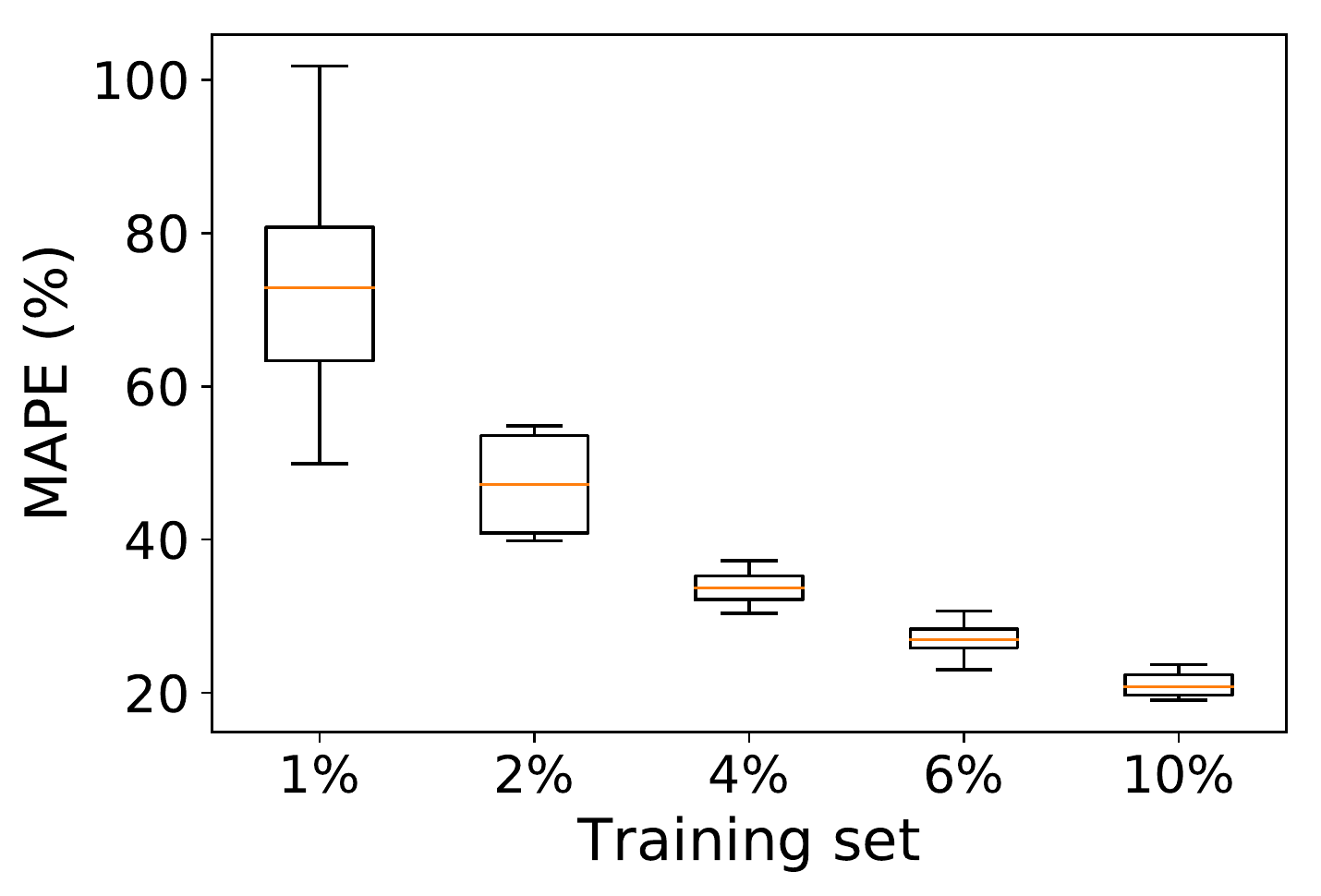}}
\end{minipage} \par\medskip
\caption*{\small (A) Stencil Computation}
\par\medskip
\vfill
\begin{minipage}{1\linewidth}
\centering
\stepcounter{figure}\addtocounter{figure}{-1}  
	\subfloat[Decision Trees]{\includegraphics[width=0.3\textwidth]{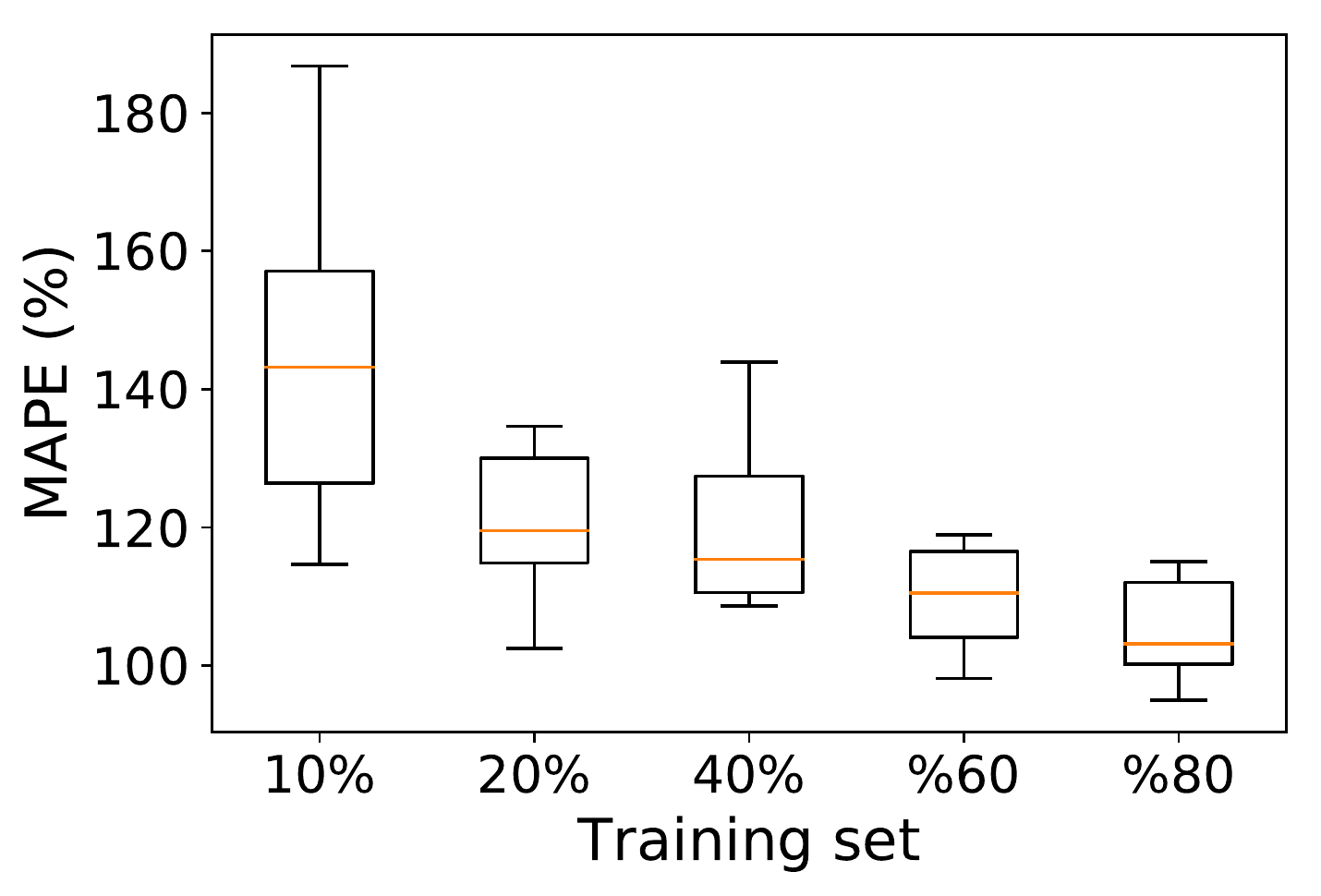}}
	\subfloat[Extra Trees]{\includegraphics[width=0.3\textwidth]{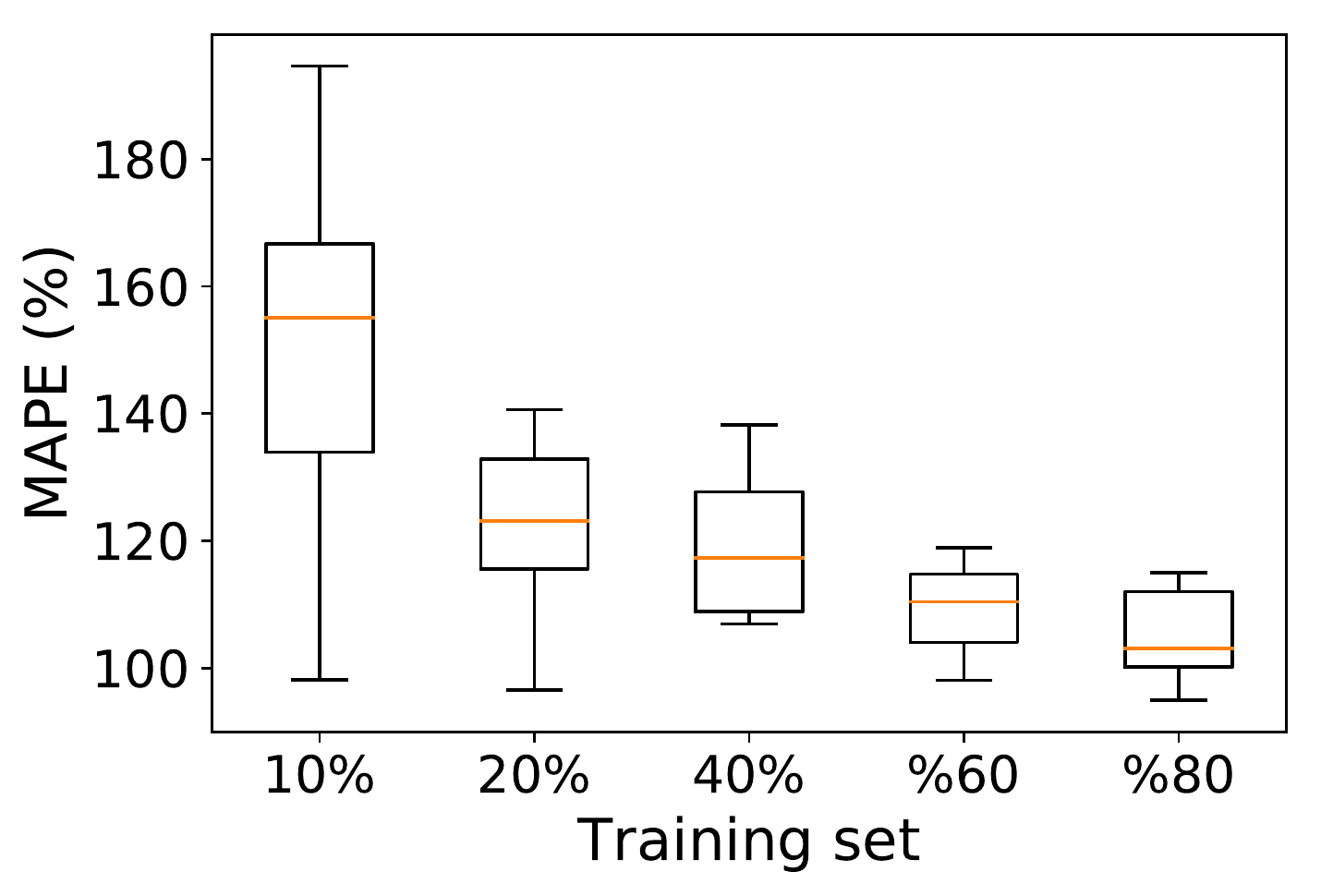}}
	\subfloat[Random Forests]{\includegraphics[width=0.3\textwidth]{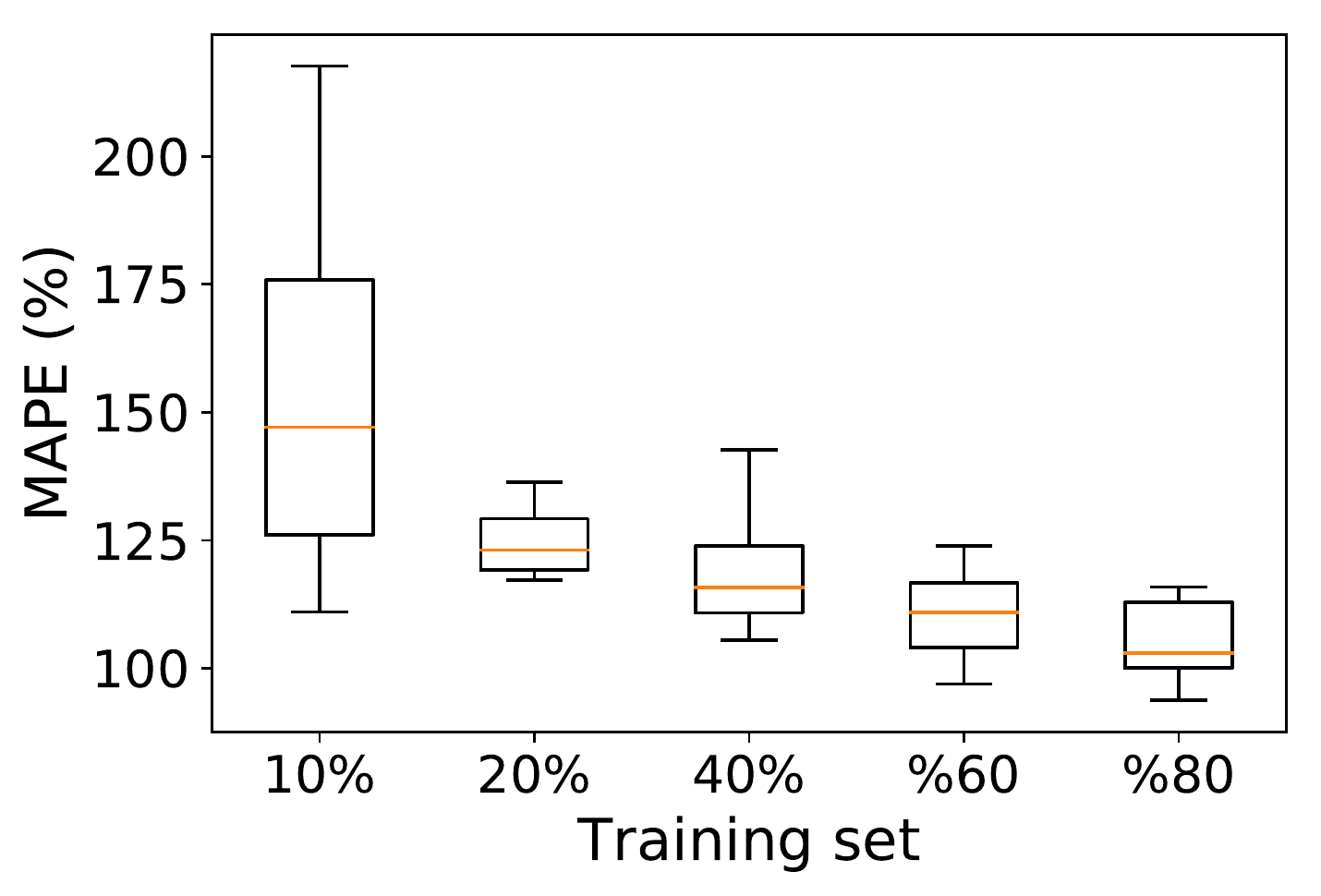}}
\end{minipage}\par\medskip
\caption*{\small (B) Fast Multipole Method}
\caption{Comparison of Mean Absolute Percentage Error ($MAPE$) scores of different machine learning models.}
\label{fig:ml}
\end{figure*}

Supervised machine learning methods have found numerous applications in performance modeling and evaluation. The basic premise of the machine learning approach is to exploit the dependencies between independent variables (feature vector) in empirical application data, while modeling the relationship between them and a response variable (execution time in this work). 

In this section, we use a suite of models from the scikit-learn package~\cite{pedregosa2011}, characterized by various degrees of complexity. In particular, we use decision trees and a set of ensemble methods including random forests and extremely randomized trees (extra trees). We employ uniform random sampling to construct the training dataset. In addition, we apply preprocessing transformation to a standard Gaussian distribution with zero mean and unit variance. Standardization of datasets is a common requirement for many machine learning estimators implemented in scikit-learn.

We test the machine learning models on two independent applications. The first is a stencil code that exposes different grid sizes and loop blocking, $\bm{X} = (I, J, K, b_i, b_j, b_k)$ where $I \times J \times K =\{1 \times 16 \times 16 \cdots 1 \times 128 \times 128\}$ with a 16 points stride and $b_i \times b_j \times b_k = \{1 \times 1 \times 1 \cdots I \times J \times K\}$. The second application is a fast multipole method code that exposes different problem sizes, particles per leaf cell, orders of expansion, and OpenMP parallelism, $\bm{X} = (t, N, q, k)$ where $t = \{1 \cdots 16\}$, $N = \{4096, 8192, 16384\}$, and $k = \{2 \cdots 12\}$. We encode information about the applications input sizes and tuning parameters into feature vectors and use the execution time as the response variable. The results of these experiments are shown in Figure~\ref{fig:ml}. We observe that the prediction accuracy improves and mean absolute percentage error ($MAPE$) variance reduces as the training dataset size increases. However, the performance of all models at small sample sizes is not satisfactory which limits their ability to model large design spaces or runtime hardware changes. Comparing different models, Figure~\ref{fig:ml} shows that extra trees model is the best performing. Thus, we use extra trees model in our hybrid approach.

\section{Approach: Hybrid Modeling}
\label{sec:hybrid}

\begin{figure}
  \centering
    \includegraphics[width=.6\linewidth]{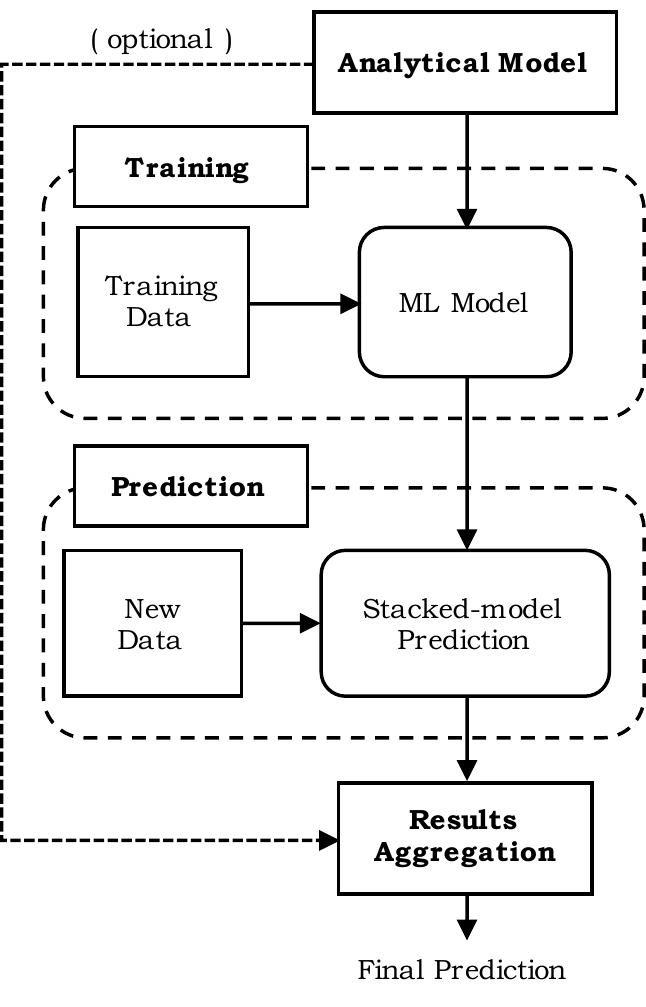}
  \caption{Hybrid performance prediction framework.}
  \label{fig:hybrid}
\end{figure}

Our proposed hybrid model consists of an analytical model, two ensemble methods, a training algorithm, and a prediction algorithm, as depicted in Figure~\ref{fig:hybrid}. Ensemble methods are meta-algorithms that combine several base models into one predictive model in order to decrease variance, bias, or improve predictions. Ensemble methods were originally developed to operate with machine learning models. We adapt these methods to support the joint usage of analytical and machine learning models. In particular, our proposed hybrid model uses stacking and bagging ensemble methods. Stacking is a way to ensemble multiple models in order to improve predictions. In stacking, the output of one model is used as input for the next level model. Bagging, on the other hand, is a method for generating multiple versions of a predictor and using these to get an aggregated prediction~\cite{wolpert1992}. Bagging reduces variance and helps to prevent overfitting~\cite{breiman1996}.

The first component of the hybrid model is an analytical model of the corresponding application code. As shown Section~\ref{sec:evaluation}, the analytical models are not required to be very accurate and need to only roughly capture the behavior of the underlying application. The second component is a training algorithm which uses a training dataset consisting of response variables and feature vectors to train a machine learning model. These feature vectors represent the modeling parameters. In addition, the machine learning model is stacked on top of the analytical model in order to improve the prediction accuracy. In other words, the analytical model predictions are regarded as additional features for the machine learning model.

Once the stacked model is constructed, it can be used to predict the performance of new data, from outside the training dataset. To make a prediction, the feature vector is passed to the stacked model, which will output a predicted response variable. For each application, the model is constructed once offline but used many times. It is not necessary to gather a training dataset or rebuild the model for every prediction.

In the last part of the hybrid model, bagging ensemble model is used to aggregate the response variables predicted by both the analytical and the stacked models. Aggregating predictions is a good technique to decrease the underlying models variance and thus reducing overfitting. This part of the hybrid model is supplementary and its benefits depend on how representative the analytical models are of the underlying application code.

\section{Evaluation}
\label{sec:evaluation}

\subsection{Effect of the Analytical Model Accuracy}

We start by evaluating the effect of the analytical model accuracy on the hybrid model prediction. To do this, we compare the performance of the hybrid approach with a pure machine learning model using a 7-point 3-D stencil code from PATUS. We use extra trees models in both approaches and compare the performance using three separate datasets with different feature vectors that the analytical models capture with various levels of accuracy.

\begin{figure}
\centering
\subfloat[Extra Trees]{\includegraphics[width=0.5\linewidth]{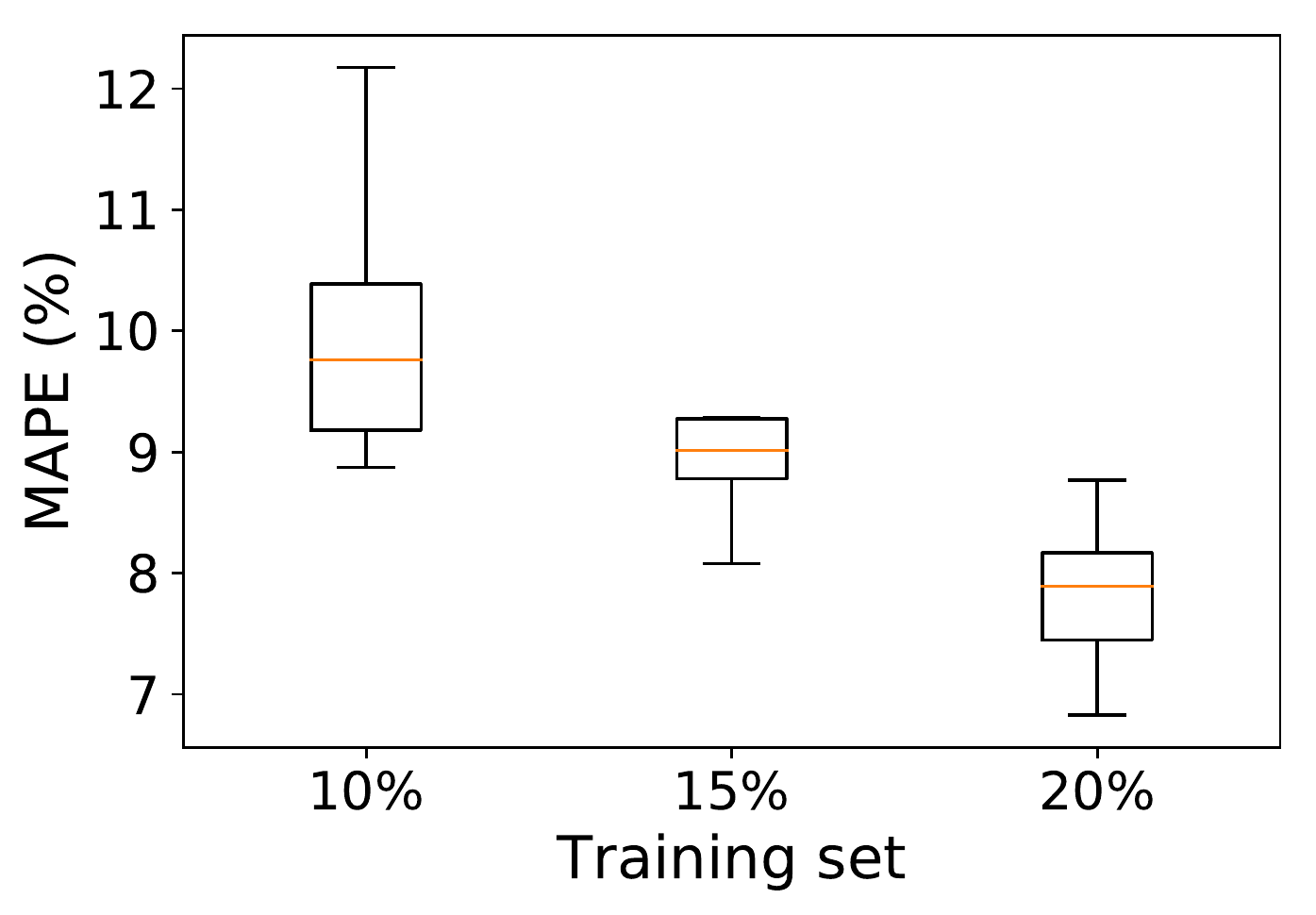}}
\subfloat[Hybrid Model]{\includegraphics[width=0.5\linewidth]{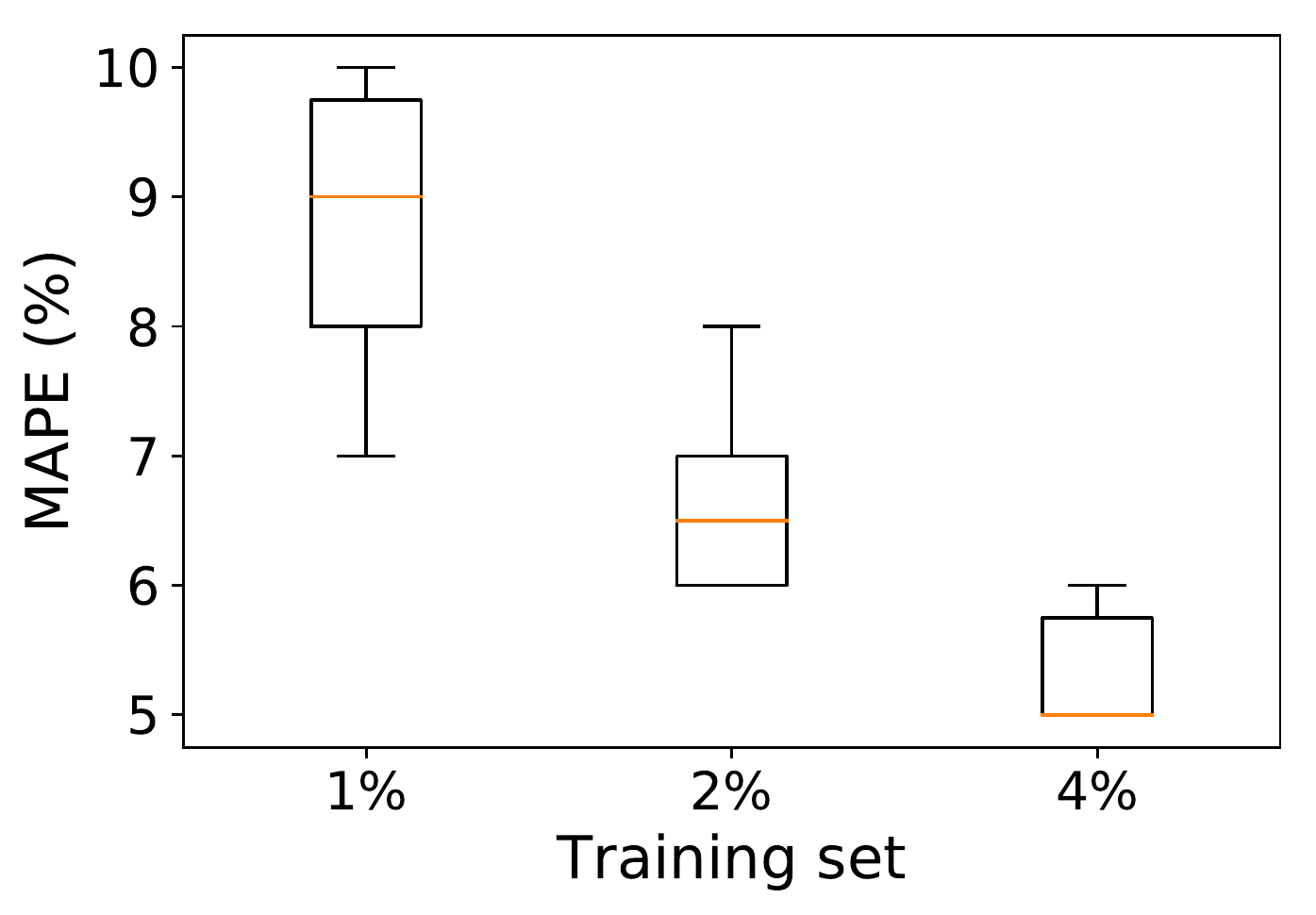}}\\
\caption{Comparison of $MAPE$ scores with various window size of the training set for stencil computation with different grid
sizes.}
\label{fig:eval_1}
\end{figure}

First, we evaluate the hybrid approach on areas that the analytical models cover accurately. Figure~\ref{fig:eval_1} shows $MAPE$ scores with various window size of the training set for the problem with different grid sizes, $\bm{X} = (I, J, K)$ where $I \times J \times K = \{128 \times 128 \times 128 \cdots 256 \times 256 \times 256\}$ with a 16 points stride. The hybrid approach reduces the size of the training set that is required to achieve $MAPE \lessapprox 10$. Here, we use $10\%$, $15\%$, and $20\%$ of the overall dataset for training in the pure machine learning approach and $1\%$, $2\%$, and $4\%$ in the hybrid approach.

Next, we add loop blocking to the analytical models. In spatial blocking, the problem domain is traversed in $TI \times TJ \times TK$ blocks. Hence, the number of blocks on each direction is given by
\begin{align*}
NBI &= I/TI,& NBJ &= J/TJ,& NBK &= K/TK.
\end{align*}

Therefore, the total number of tiling iterations to perform is $NB = NBI \times NBJ \times NBK$. Spatial blocking is considered into the analytical models introduced in Section~\ref{sec:AM} by reassigning $I$, $J$, and $K$ and their extended dimensions as follows
\begin{align*}
I &= \ceil[\bigg]{\frac{TI}{W}} \times W, & II &= \ceil[\bigg]{\frac{TI + 2 \times l}{W}} \times W,\\
J &= TJ, & JJ &= TJ + 2 \times l,\\
K &= TK, & KK &= TK + 2 \times l.
\end{align*}

The new $II$, $JJ$, and $KK$ parameters are then used to estimate $nplanes_{Li}$. Finally,~\eqref{eq:misses} is rewritten as
\begin{equation}
Misses_{Li} = \lceil II / W \rceil \times JJ \times KK \times nplanes_{Li} \times NB.
\end{equation}

\begin{figure}
\centering
\subfloat[Extra Trees]{\includegraphics[width=0.5\linewidth]{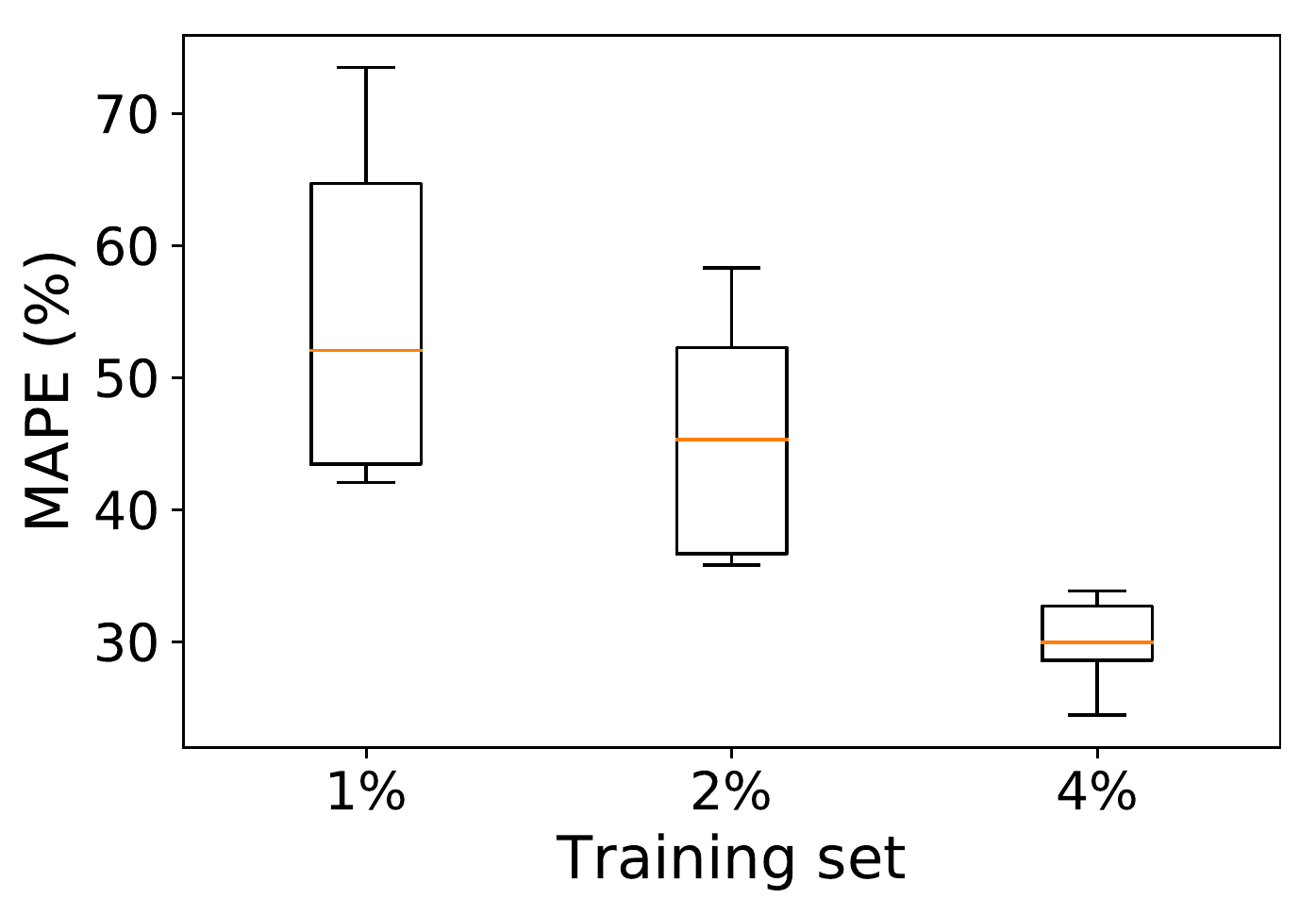}}
\subfloat[Hybrid Model]{\includegraphics[width=0.5\linewidth]{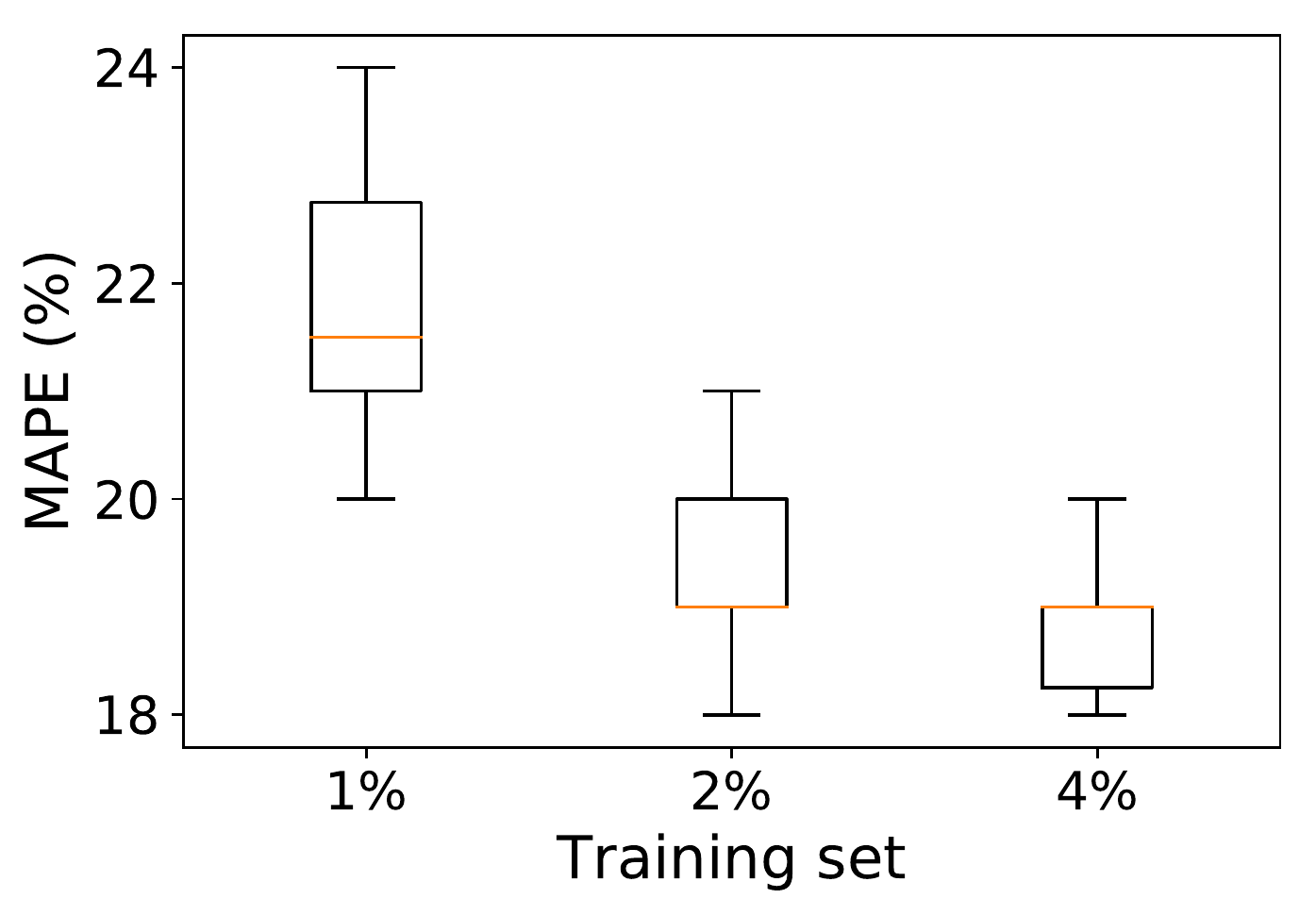}}\\
\caption{Comparison of $MAPE$ scores with various window size of the training set for stencil computation with different grid sizes and loop blocking.}
\label{fig:eval_2}
\end{figure}

The cache access time of the blocked code is implementation dependent. Therefore, analytical models tuning is needed to achieve sufficient accuracy. However, we do not tune the analytical models as our goal here is to study the effect of using inaccurate analytical models on the hybrid approach (Here, analytical model $MAPE = 42\%$). Figure~\ref{fig:eval_2} shows $MAPE$ scores of the pure machine learning model and the hybrid approach using a dataset with feature vectors representing different grid sizes and loop blocking information, $\bm{X} = (I, J, K, b_i, b_j, b_k)$ where $I \times J \times K = \{1 \times 16 \times 16 \cdots 1 \times 128 \times 128\}$ with a 16 points stride and $b_i \times b_j \times b_k = \{1 \times 1 \times 1 \cdots I \times J \times K\}$. As shown in Figure~\ref{fig:eval_2}, incorporating the analytical models cuts the percentage error nearly in half.

\begin{figure}
\centering
\subfloat[Extra Trees]{\includegraphics[width=0.5\linewidth]{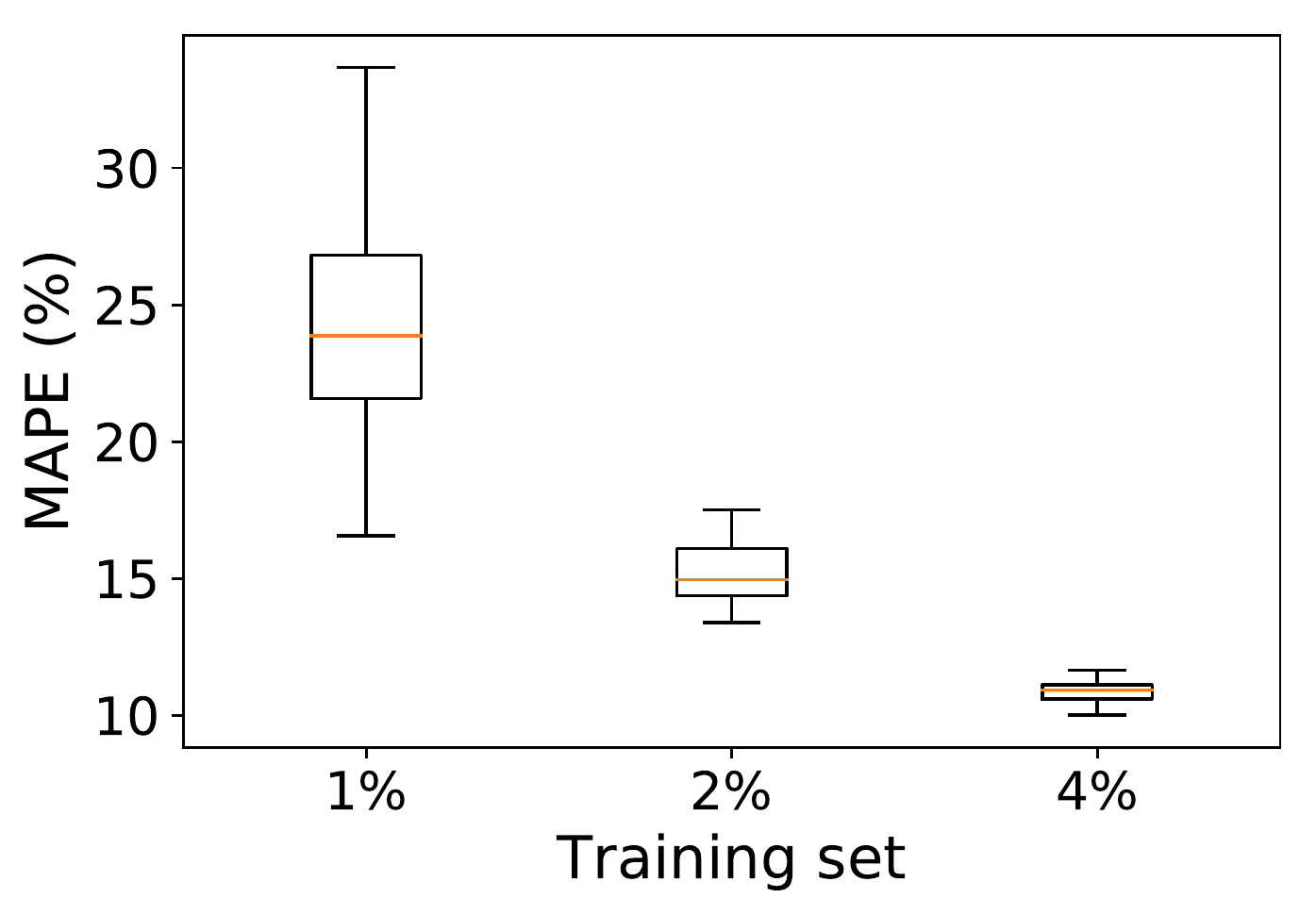}}
\subfloat[Hybrid Model]{\includegraphics[width=0.5\linewidth]{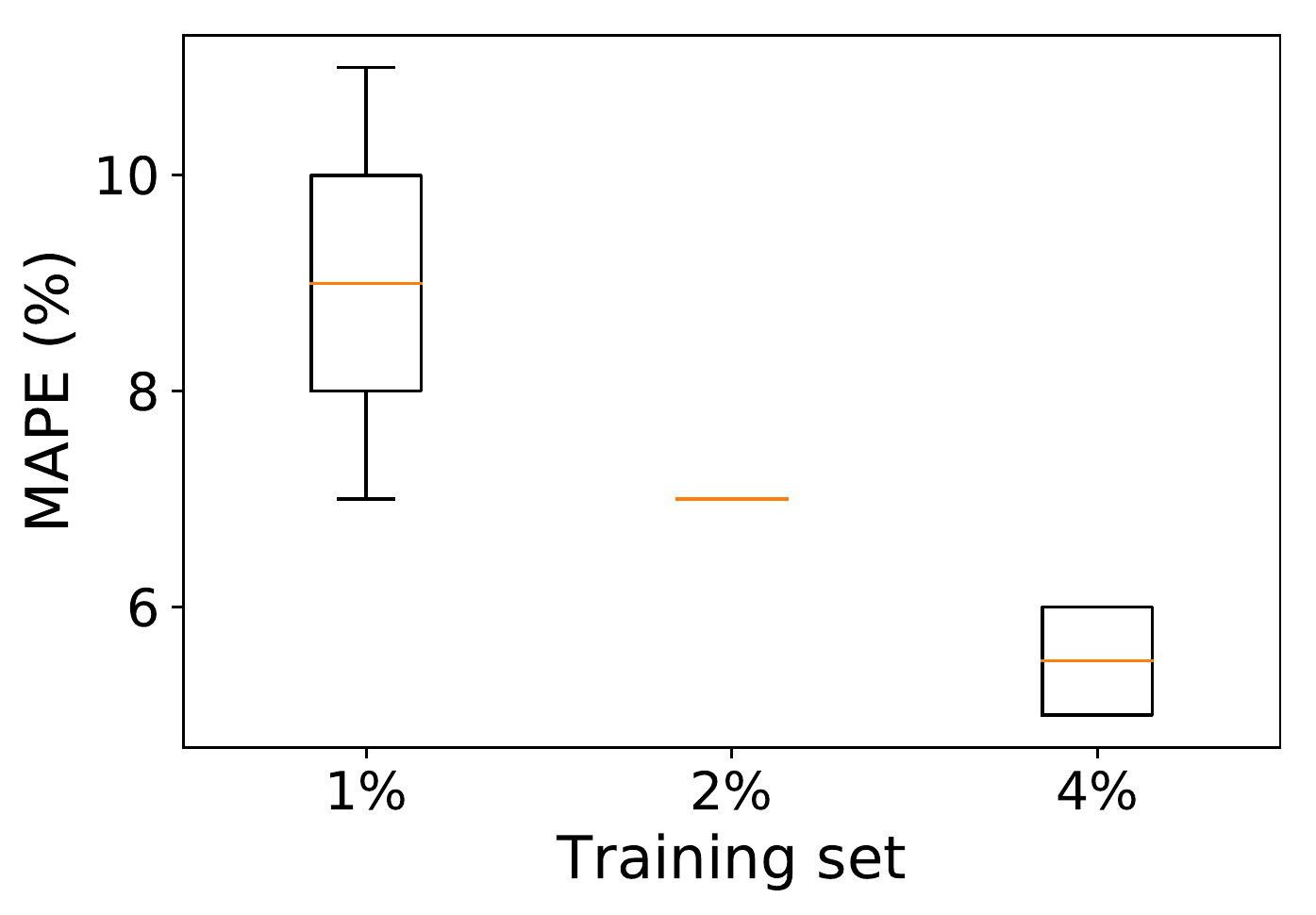}}\\
\caption{Comparison of $MAPE$ scores with various window size of the training set for stencil computation with different grid sizes and multithreaded parallelism.}
\label{fig:eval_3}
\end{figure}

Lastly, we evaluate the hybrid model on a region that is not covered by the analytical models. In particular, the analytical models presented in Section~\ref{sec:AM} are designed to model the performance on a single core and do not cover multithreaded parallelism. We couple these analytical models with extra trees machine learning model and evaluate the hybrid approach using a dataset obtained by running a multithreaded stencil code with different grid sizes, $\bm{X} = (I, J, K, t)$ where $I \times J \times K = \{128 \times 128 \times 1 \cdots 176 \times 176 \times 1\}$ with a 16 points stride and the number of threads $t = \{1 \cdots 8\}$. Here we do not aggregate the analytical and stacked models predictions as the analytical models do not capture the parallelism. Figure~\ref{fig:eval_3} shows that using serial analytical models in the hybrid approach can improve the prediction accuracy for the parallel dataset in comparison to the pure machine learning model.

\subsection{Tuning Application Parameters}

\begin{figure}
\centering
\subfloat[Extra Trees]{\includegraphics[width=0.5\linewidth]{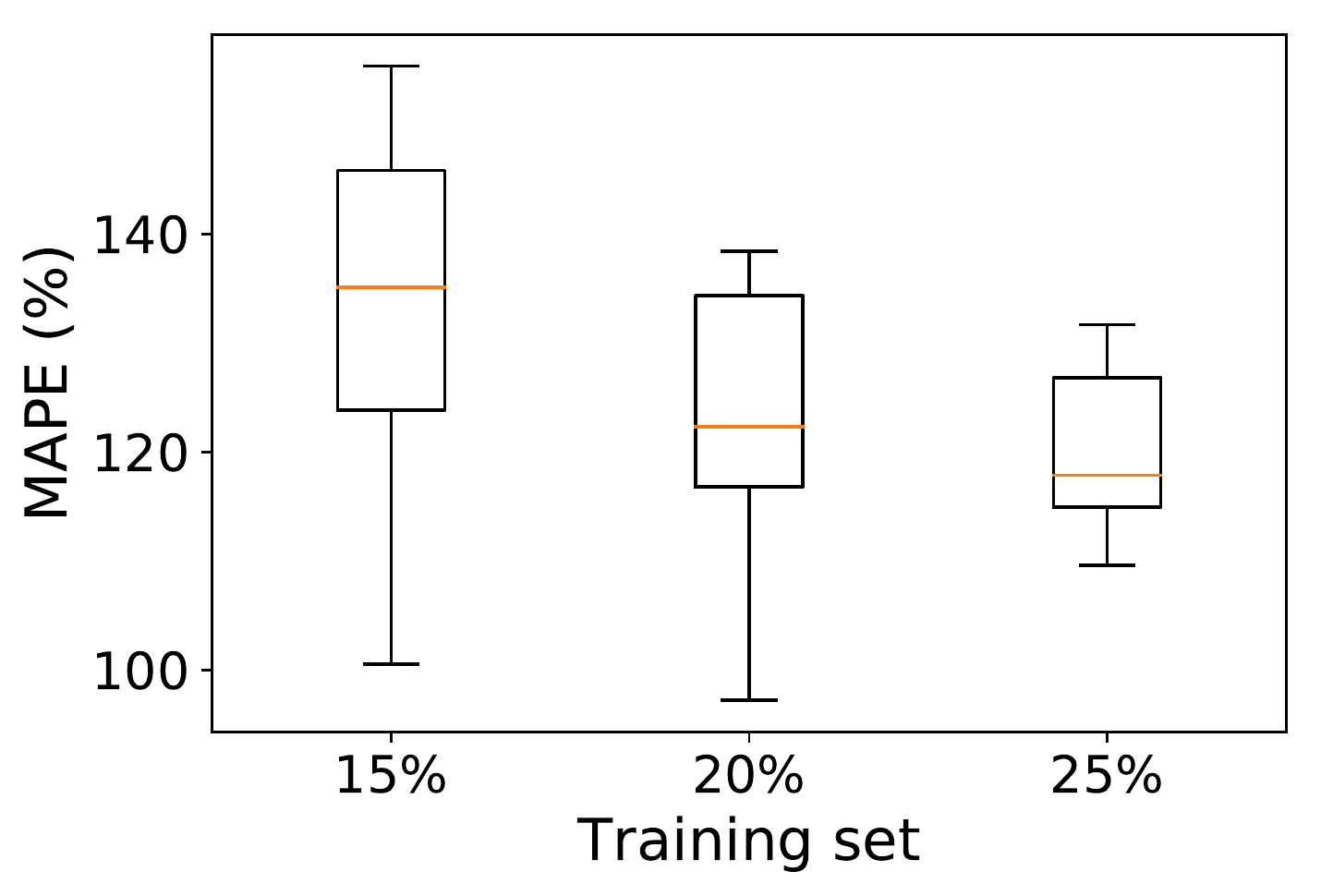}}
\subfloat[Hybrid Model]{\includegraphics[width=0.5\linewidth]{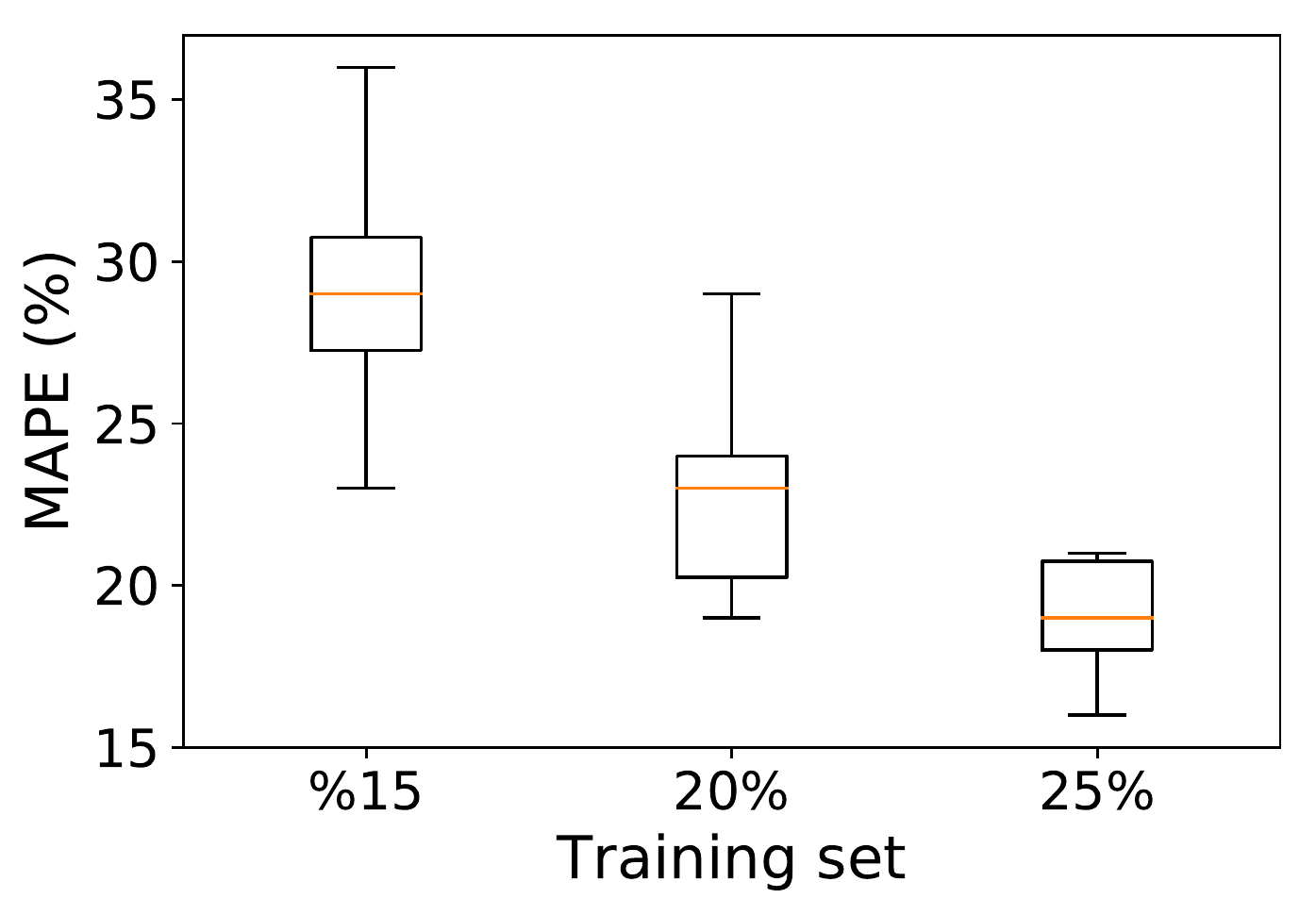}}\\
\caption{Comparison of $MAPE$ scores with various window size of the training set for fast multipole method.}
\label{fig:eval_F}
\end{figure}

Here we use the hybrid approach to model the performance of the FMM using feature vectors consisting of four components $\bm{X} = (t, N, q, k)$ where $t = \{1 \cdots 16\}$ is the number of threads, $N = \{4096, 8192,16384\}$ is the number of particles, $q$ is the number of particles per leaf cell, and $k = \{2 \cdots 12 \}$ is the order of expansion. Fast multipole method is a highly complex algorithm with several different phases, a combination of data structures, fast transforms, and irregular data access. As shown in Figure~\ref{fig:ml}, pure machine learning models are not able to accurately predict the FMM performance even with a very large training dataset. 

We couple the FMM analytical models presented in Section~\ref{sec:AM} with extra trees machine learning model. Again, we do not tune the analytical models relative to the underlaying implementation (analytical model $MAPE = 84.5\%$). We compare the accuracy of the hybrid approach with the pure machine learning model. Figure~\ref{fig:eval_F} shows that the hybrid model is able to significantly improve the $MAPE$ scores of the pure machine learning model. However, because of the high complexity of the FMM algorithm, the FMM hybrid model requires larger training dataset to carry out predictions with reasonable accuracy in comparison to stencil computation.

\section{Related Work}
\label{sec:related}

Using analytical models for performance modeling and prediction is a broadly searched topic. Hoefler \etal~\cite{Hoefler2010} describes the importance of analytical performance models for parallel applications to understand the performance implications of different choices of algorithms and implementation options. Models that predict performance of stencil computations are used in~\cite{de2011,de2015}. Performance models for the computations and communication in fast multipole methods are presented in~\cite{Chandramowlishwaran2012} and~\cite{Yokota2014,Ibeid2016}, respectively.

There is also plenty of research conducted on using machine learning for performance modeling. Sun \etal~\cite{sun2017} focuses on automatically predicting the execution time of parallel program by collecting data from executions with different inputs. Lee \etal~\cite{Lee2007} builds and measures regression and neural network models for understanding large application parameter spaces. Marathe \etal~\cite{Marathe2017} utilizes transfer learning to select the best performing configurations at a target large-scale scenario using domain knowledge extracted from a source small-scale scenario. The use of machine learning to predict the performance of stencil kernels was considered in~\cite{martinez2017, cosenza2017}.

While analytical modeling and machine learning techniques are two traditional methods for performance modeling, combining the two is a new approach that has been receiving increasing attention in recent years. Didona \etal~\cite{Didona2015} explores several ways to combine analytical and machine learning models and evaluate the proposed techniques on two middleware systems, a NoSQL distributed key-value store and a Total Order Broadcast (TOB) service. Ehsan \etal~\cite{ataie2016} uses synthetic data generated by analytical models to train a machine learning model for performance prediction of Map-Reduce jobs in cloud environment. Didona \etal~\cite{didona2014_1, didona2014_2} use the divide and conquer technique to model the performance of distributed transactional applications. This technique consists of building performance models for individual parts of the entire system based on either analytical or machine learning models.

\section{Conclusion}
\label{sec:conclusion}

In this paper, we propose and validate a hybrid approach for performance modeling that couples analytical and machine learning techniques in order to take advantage of both methods. The hybrid model uses stacking and bagging ensemble methods to improve predictions, reduce variance, and avoid overfitting. Our results show that the hybrid approach is effective in predicting the execution time by reducing the $MAPE$ score of pure machine learning models. In addition, the hybrid model requires small training dataset to carry out predictions with reasonable accuracy, thus making it suitable for hardware and workload changes.

\section*{Acknowledgment}

The authors would like to thank Prof. David Keyes (KAUST) for many insightful discussions. This material is based in part upon work supported by the Department of Energy, National Nuclear Security Administration, under Award Number DE-NA0002374.

\bibliographystyle{IEEEtran}
\bibliography{ref}

\end{document}